\documentclass[twocolumn,aps,showpacs,preprintnumbers,lettersize]{revtex4}
\usepackage{amssymb}
\usepackage{graphicx}
\usepackage{dcolumn}
\usepackage{bm}
\usepackage{color}

\def\ga{{\ \lower-1.2pt\vbox{\hbox{\rlap{$
          >$}\lower5pt\vbox{\hbox{$\sim$}}}}\ }}
\def\la{{\ \lower-1.2pt\vbox{\hbox{\rlap{$
          <$}\lower5pt\vbox{\hbox{$\sim$}}}}\ }}
\def\beq{\begin{equation}}
\def\eeq{\end{equation}}
\def\bea{\begin{eqnarray}}
\def\eea{\end{eqnarray}}
\def\he3he4{$^3$He-$^4$He}
\def\eps{\varepsilon}

\def\cost2e{\cos\frac{\theta}{2}}
\def\cost2t{\cos\theta/2}
\def\he3{$^3$He}
\def\he4{$^4$He}

\def\g1t{$\tilde{\Gamma}_1$}

\definecolor{og}{named}{OliveGreen}

\begin{document}

\title{Superconducting fluctuation corrections to ultrasound attenuation in
layered superconductors}
\author{M.S. Mar'enko$^{1}$}
\author{C. Bourbonnais$^{1,2}$}
\author{A.-M.~S.~Tremblay$^{1,2}$}
\affiliation{$^{1}$D\'{e}partement de physique and Regroupement qu\'{e}b\'{e}cois sur les
mat\'{e}riaux de pointe, Universit\'{e} de Sherbrooke, Sherbrooke, Qu\'{e}%
bec, J1K 2R1, Canada\\
$^{2}$Institut canadien de recherches avanc\'{e}es, Universit\'{e} de
Sherbrooke, Sherbrooke, Qu\'{e}bec, J1K 2R1, Canada }
\date{\today }

\begin{abstract}
We consider the temperature dependence of the sound attenuation and sound
velocity in layered impure metals due to $s$-wave superconducting fluctuations of the
order parameter above the critical temperature. We obtain the dependence on
material properties of these fluctuation corrections in the hydrodynamic
limit, where the electron mean free path $\ell $ is much smaller than the
wavelength of sound and where the electron collision rate $\tau ^{-1}$ is
much larger than the sound frequency. For longitudinal sound propagating
perpendicular to the layers, the open Fermi surface condition leads to a
suppression of the divergent contributions to leading order, in contrast
with the case of paraconductivity. The leading temperature dependent
corrections, given by the Aslamazov-Larkin, Maki-Thompson and density of
states terms, remain finite as $T\rightarrow T_{c}$. Nevertheless, the
sensitivity of new ultrasonic experiments on layered organic conductors
should make these fluctuations effects measurable.
\end{abstract}

\pacs{74.25.Ld, 74.25.Fy, 74.40.+k}
\maketitle



The problem of fluctuations of the superconducting order parameter above the
critical temperature began to attract the attention of researchers more than
three decades ago.\ \cite{AL,Maki,Thompson} Considerable work has been done
from both the theoretical and experimental points of view. Conductivity,
thermoconductivity, magnetoconductivity as well as tunneling properties and
nuclear magnetic resonance characteristics are amongst the various transport
phenomena where fluctuation contributions were predicted and experimentally
observed. Ref.\ \cite{LV01} contains an extensive review of the field and
references.

In the present paper, we concentrate on the superconducting fluctuation
corrections to the sound velocity and sound attenuation in layered
superconductors above $T_{c}$. As usual, superconducting fluctuations
manifest themselves in three different ways.\ \cite{LV01} a) The effective
number of normal carriers is reduced because some of the electrons exist as
transient Cooper pairs. This is the so-called density of states
contribution. b) The single-particle excitations are Andreev reflected off
the superconducting fluctuations, as described by the so-called
Maki-Thompson term. c) Some of the electrons behave like Cooper pairs for a
time given by the Ginzburg-Landau time. This is the famous Aslamazov-Larkin
contribution. Layered organic superconductors, in particular, are an ideal
class of materials to study fluctuation phenomena. This is because the
relatively low charge-carrier concentration and the strong electronic
anisotropy enhance the effect of superconducting fluctuations and increase
the size of the fluctuation regime. The manifestation of superconducting
fluctuations in organic materials was found in a number of experimental
works on dc-magnetoconductivity, thermoconductivity, ac-susceptibility,
specific heat and torque magnetometry.\ \cite{fluct_ex} Preliminary
ultrasonic experiments in the $\kappa$-phase layered organic materials \cite%
{usherb} (BEDT-TTF)$_{2}$Cu[N(CN)$_{2}$]Br, (BEDT-TTF)$_{2}$Cu[N(CN)$_{2}$%
]Cl and (BEDT-TTF)$_{2}$Cu(NCS)$_{2}$ suggest that the sound attenuation
coefficient is decreased by superconducting fluctuations over a sizeable
temperature range above $T_{c}$. However, the effect is much weaker than the
fluctuation correction to the conductivity\ \cite{KFrikash} where, in the
same compounds, the Aslamazov-Larkin contribution dominates the
superconducting fluctuation effects. Although sound attenuation experiments
seem to be a very promising method to study fluctuation phenomena in layered
superconductors, the theoretical description is still incomplete and needs
to be analyzed in more details. This more detailed analysis can tell us, for
example, which material properties control fluctuation phenomena.
Fluctuation phenomena can then be used to access these properties when they
are not independently known. For example, we will see that the phase
breaking time can be very important to determine the fluctuation
contribution to ultrasonic attenuation whereas it would not contribute to
dominant terms in the corresponding fluctuation conductivity. It is by
considering a large set of experimental probes that one can make sure that
all material properties are unambiguously determined.

In this paper we assume, as usual for fluctuation corrections, that we are
close enough to $T_{c}$ that the characteristic pair frequency is less than
temperature (in dimensionless units). This is the so-called
renormalized-classical regime. We study the case where longitudinal sound
propagates perpendicular to the layers. This is the most relevant case to
study since it is most easily accessible experimentally.\ \cite{KFrikash} We
also consider the hydrodynamic limit where the electron mean free path $\ell
$ is much smaller than the wavelength of sound and where the electron
collision rate $\tau ^{-1}$ is much larger than the sound frequency. In
quasi two-dimensional systems, the Fermi surface is opened in the direction
perpendicular to the layers. The coupling to the sound comes from the
modulation of the hopping integral across the layers, taking also into
account the electroneutrality condition in the moving frame.\ \cite%
{SamokhinWalker} The Fr\"{o}hlich model where phonons couple to the
electronic density is not valid. {\ The most dramatic consequence of the
open Fermi surface in such a situation is the suppression of the divergence
in the Aslamazov-Larkin and Maki-Thompson terms that often give, by
contrast, the most important contribution to the fluctuation correction to
the electrical conductivity.} Hence, sound propagation is less strongly
modified by superconducting fluctuations than the electrical conductivity.
All the calculations are performed for $s-$wave superconducting
fluctuations. The $d-$wave case needs a separate treatment.\ \cite{d-wave,Sauls_tau}

In the following section, we discuss earlier work and present the
Hamiltonian. The diagrams that enter the calculation and then details of the
calculation appear in the following sections. The results are gathered in
Sec.\ \ref{Sec.Results} and discussed in Sec.\ref{Sec_Discussion}. The
appendix contains mathematical details on the analytic continuation of
digamma functions.

\section{The model and definitions}

Ultrasonic attenuation in impure metals was analyzed by Pippard \cite%
{Pippard}, using the Boltzmann equation method, and by Tsuneto \cite{Tsuneto}
with the density matrix formalism. Later on, Schmid \cite{Schmid}, using
Green function methods, derived the effective electron-phonon interaction in
the presence of impurities taking into account the screening of Coulomb
repulsion between electrons. A key point of the approach of Tsuneto, Pippard
and Schmid was to consider the electron system in a reference frame moving
together with the ion lattice. In an impure metal, the elastic scattering of
electrons as well as perfect screening at small phonon momentum and
relaxation to equilibrium occur in this oscillating frame. In this
formalism, the electron-phonon coupling appears through the stress tensor
instead of through the density operator occurring in the Fr\"{o}hlich model.
The latter model leads to erroneous results in the dirty limit.\ \cite%
{Schmid}

The idea of a moving reference frame was also used by Kotliar and
Ramakrishnan \cite{Kotliar}, who considered the limit of strongly disordered
metal and analyzed the effect on ultrasound propagation of incipient
localization near the metal-insulator transition. Reizer \cite{Reizer}
considered the effect of various types of inelastic scattering (including
electron-electron, electron-magnon scattering and weak-localization effects)
on sound attenuation.

The fluctuation corrections to the ultrasound attenuation in 3D metals were
estimated in the most singular channel in the early work of Aslamazov and
Larkin.\ \cite{AL} Aslamazov and Varlamov, back in 1979, found the
fluctuation corrections in layered superconductors.\ \cite{AV} However,
these papers did not consider explicitly the form of the density of states
for a corrugated cylindrical Fermi surface. That density of states is energy
independent contrary to the assumption made in these papers. In addition,
the calculation was done for the Fr\"{o}hlich model. Also, Ref.\ \cite{AV}
used impurity vertex corrections for the external vertices in the
correlators and we shall see that these are unnecessary in the hydrodynamic
limit when one takes into account the fact that the calculation of impurity
scattering should be done in the electrically neutral moving frame.\ \cite{Schmid,Kotliar} For the
case of interest, with open Fermi surface in the perpendicular direction, we find results that are
quite different from these earlier works.

We use the unperturbed quasiparticle energy spectrum
\begin{equation}
\xi (p_{||},p_{z})=\frac{p_{||}^{2}-p_{F}^{2}}{2m}+2t_{\perp }\cos {(p_{z}a)}%
,  \label{spectrum}
\end{equation}%
appropriate for layered metals. Here, $p_{z}$ is the component of the
quasiparticle momentum perpendicular to the conducting planes, $p_{||}$ is
the in-plane momentum, $t_{\perp }\ll \varepsilon _{F}$ is the small
interlayer hopping energy, and $a$ is the interlayer distance. In this
model, the Fermi surface has the well-known shape of a corrugated cylinder,
and the density of states is given by $\nu _{0}=m/(2\pi a)$. It is important
to note that in this model the density of states $\nu _{0}$ does not depend
on the quasiparticle energy. In other words, the derivative $(\partial \nu
_{0}/\partial \xi )\equiv 0$ vanishes. This relation is obviously true at $%
t_{\perp }=0$, i.e. for non-interacting two-dimensional layers. If there is
a small interlayer interaction $t_{\perp }\neq 0$, any cross-section of the
Fermi surface made at any value of $p_{z}$ still gives the two-dimensional
(2D) result with $(\partial \nu _{0}/\partial \xi )\equiv 0$. One can easily
see that this relation is valid for any Fermi surface that is opened in the
third direction ($t_{\perp }\ll \varepsilon _{F}$) and has circular cross
section in the plane.

As shown by Tsuneto \cite{Tsuneto} and Schmid \cite{Schmid}, based on
earlier work of Pippard \cite{Pippard} and Blount, \cite{Blount} the
canonical transformation that takes us to the moving frame leads, in the
continuum limit, to an electron-phonon interaction that originates from the
commutator of the infinitesimal generator of the transformation with the
kinetic energy operator. The remaining contributions from the commutator of
the generator of the canonical transformation with the electron-ion and
electron-electron interactions are overall negligible. Since a trace over
fermions is performed to compute phonon damping, the result should be
independent of the canonical transformation. The moving frame is the one
best suited for approximations.

For a tight binding model, such as the one we need to consider to obtain the
proper open Fermi surface for propagation in the perpendicular direction $%
p_{z}$, Eq.\ (\ref{spectrum}), the corresponding microscopic derivation of
the stress tensor has not been performed yet. It is not our purpose to give
this derivation here. Instead, we follow the footsteps of Barisic,\ \cite%
{Barisic} Su-Schrieffer-Heeger, \cite{SuSchrieffer} and, more recently,
Walker, Smith, and Samokhin\ \cite{SamokhinWalker} and assume that the
electron-phonon interaction comes from the modulation of the hopping
parameter $t_{\perp }$ induced by the lattice deformation. Neglecting
umklapp processes, one obtains for the interaction of a longitudinal phonon
propagating in the $z$ direction perpendicular to the planes

\begin{widetext}
\begin{equation}
H_{e-ph}=-\sqrt{2}iG\sum_{\mathbf{k,p}}\left( \frac{\hbar \omega _{0}\left(
\mathbf{k}\right) }{NMv_{s}^{2}}\right) ^{1/2}\left( \widehat{\mathbf{k}}
\mathbf{\cdot }\widehat{\mathbf{z}}\right) \left( \widehat{\mathbf{e}}
\mathbf{\cdot }\widehat{\mathbf{z}}\right) \left( \cos p_{z}a\right) \text{ }
c_{\mathbf{p}+\mathbf{k},\sigma }^{\dagger }c_{\mathbf{p,}\sigma }\left(
a_{- \mathbf{k}}^{\dagger }+a_{\mathbf{k}}\right) ,  \label{He-ph}
\end{equation}
\end{widetext}%
where $\omega _{0}\left( \mathbf{k}\right) =v_{s}k$ is the sound frequency, $%
v_{s}$ the sound velocity, $M$ the ion mass, $N$ the number of unit cells, $%
G $ a constant that depends on the derivative of the hopping integral $%
t_{\perp }$ with respect to the strain, while $a_{\mathbf{k}}^{\left(
\dagger \right) },c_{\mathbf{p},\sigma }^{\left( \dagger \right) }$ are,
respectively, destruction and creation operators for phonons and for
electrons of spin $\sigma $. The above expression is restricted to
longitudinal phonons propagating in the perpendicular direction, hence the
wave vector and polarization of the phonons involved in the interaction will
satisfy $\widehat{\mathbf{k}}\mathbf{\cdot }\widehat{\mathbf{z}}=\widehat{%
\mathbf{e}}\mathbf{\cdot }\widehat{\mathbf{z}}=1$. The above expression
neglects the compression and stretching of chemical bonds that are not
strictly along the direction of propagation.

\section{Sound attenuation without fluctuation correction\label{Normal}}

\begin{figure}[h]
\includegraphics[width=7.0cm]{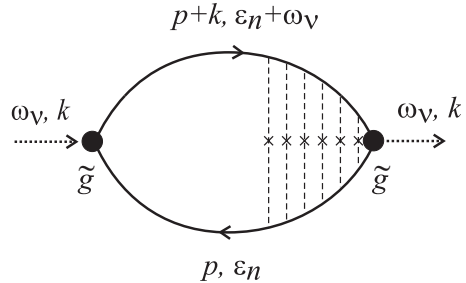}
\caption{The phonon self-energy in the case of an impure metal. Solid lines
are one-electron Green functions, dotted lines the bare phonon propagators,
dashed lines with crosses correspond to the impurity scattering, and bold
circles show the renormalized electron-phonon vertices.}
\label{poloper}
\end{figure}

The sound attenuation coefficient is determined by the imaginary part $
\gamma (k)$ of the complex frequency $\omega (k)$ {where }the pole of the
phonon Green function $D(\mathbf{k},\omega )$ {is located}. This quantity
obeys Dyson's equation \cite{AGD}
\begin{equation}
D^{-1}(\mathbf{k},\omega _{\nu })=\left( D^{0}(\mathbf{k},\omega _{\nu
})\right) ^{-1}-\Pi (\mathbf{k},\omega _{\nu }).  \label{Dyson_eq}
\end{equation}
Expressed in bosonic Matsubara frequencies $\omega _{\nu }=2\pi \nu T$ using
units $k_{B}=1$ $\hbar =1$, with $\nu $ an integer, the quantity $D^{0}(
\mathbf{k},\omega _{\nu })$
\begin{equation}
D^{0}(\mathbf{k},\omega _{\nu })=-\frac{\omega _{0}^{2}(\mathbf{k})}{\omega
_{\nu }^{2}+\omega _{0}^{2}(\mathbf{k})}  \label{Phonon_propagator}
\end{equation}
is the phonon propagator in the non-interacting case and $\Pi (\mathbf{k}%
,\omega _{\nu })$ is the phonon self-energy. In an impure metal, the
diagrammatic representation of $\Pi (\mathbf{k},\omega _{\nu })$ is given by
Fig.\ \ref{poloper}. In the theory of electronic conductivity, one would
consider a similar diagram for the current-current correlator with the bold
dots representing the components of the vector velocity $ev_{\alpha}=e(%
\partial \xi /\partial p_{\alpha })$, where $\xi $ is the electronic energy
and $p_{\alpha }$ the $\alpha $ component of momentum. In the case of sound
attenuation, the vertex comes instead from the stress tensor corresponding
to the electron-phonon interaction Eq.\ (\ref{He-ph}). Using Eqs.\ (\ref%
{He-ph}) and (\ref{Phonon_propagator}), one obtains for this vertex $g\cos
(p_{z}a)$ where $g$ is a constant.

Each solid line in Fig.\ \ref{poloper} corresponds to the electron Green
function which, at finite temperature in the presence of impurities, is
given by
\begin{equation}
G(\mathbf{p},\varepsilon _{n})=\frac{1}{i\tilde{\varepsilon}_{n}-\xi (
\mathbf{p})},
\end{equation}
where $\xi (\mathbf{p})$ is the quasiparticle energy and where we defined $
\tilde{\varepsilon}_{n}=\varepsilon _{n}+1/(2\tau )\mathrm{sign}(\varepsilon
_{n})$ with $\varepsilon _{n}=\pi T(2n+1)$ the Matsubara frequency and $\tau
$ the elastic scattering time. The quantity $i/(2\tau )\mathrm{sign}%
(\varepsilon _{n})$ is the imaginary part of the quasiparticle self-energy.
The real part of the self-energy is constant and is absorbed in the
definition of the chemical potential.

In principle, one should include the vertex corrections that are
represented, to leading order in $(p_{F}\ell)^{-1}$ with $\ell$ the electron
mean free path, by the non-intersecting dashed impurity lines, as shown in
Fig.\ \ref{poloper}. In the case of the current-current correlator, vertex
corrections lead to the replacement of the scattering time $\tau $ by its
transport analog $\tau_{tr}$.\ \cite{AGD} For $s$-wave scattering, vertex
corrections vanish because the vector vertex $ev_{\alpha }$ averages to zero
upon angular integration at vanishing external momentum $\mathbf{k}$. This
argument generally does not work with the correlator of the stress tensor.
However, taking into account perfect screening, it was shown by Schmid \cite%
{Schmid} in the continuum limit that there is no impurity diffusion
enhancement of the electron-phonon vertex in the case of transverse phonons,
and that for longitudinal phonons this effect is negligible in the
hydrodynamic limit $k \ell \ll 1$ and $\omega \tau \mathbf{\ll }1$.
Physically, this comes about from the fact that the calculation should be
done in the moving frame and that screening is perfect at long wavelengths.
In our case, the analogous argument of electroneutrality leads \cite%
{SamokhinWalker} to the replacement of the stress vertex $F$ by $%
F-\left\langle F\right\rangle $, where the average accounts for the chemical
potential shift.\ \cite{Abrikosov} That average is precisely what is needed
to make the impurity vertex correction vanish. In addition, in our specific
case, $\left\langle F\right\rangle =\left\langle g\cos (p_{z}a)\right\rangle
=0$ to the order in phonon wave vector $k$ that we need.

The phonon self-energy without fluctuation correction may finally be
obtained from
\begin{eqnarray}
&&\Pi (k,\omega _{\nu }) \\
&&{\!\!\!\!\!\!\!\!\!\!\!\!\!}=2g^{2}T\sum_{\varepsilon _{n}}\int \frac{
d^{3}p}{(2\pi )^{3}}\cos ^{2}(p_{z}a)G(\varepsilon
_{n},p)G(\varepsilon_{n}+\omega _{\nu },p+k).  \nonumber
\end{eqnarray}
To evaluate this integral when $\tau $ is finite, it is easiest to make the
change of variables $\xi =\frac{p_{||}^{2}-p_{F}^{2}}{2m}+2t_\perp\cos
(p_{z}a)$
\begin{equation}
\int \frac{d^{3}p}{(2\pi )^{3}}=\frac{m}{(2\pi )^{3}}\int_{-\infty
}^{\infty}d\xi \int_{0}^{2\pi} d\theta \int_{-\pi /a}^{\pi /a}dp_{z}
\end{equation}
and to evaluate the integral over $\xi $ first, being careful to add and
subtract the clean limit result to insure convergence. After summation over $%
\varepsilon _{n}$ one should make the analytic continuation of the external
phonon frequency following the rule $i\omega _{\nu }\rightarrow \omega
+i\delta $. For sound, it is justified to work in the hydrodynamic limit $%
\omega \tau \ll 1$ even for very pure systems. The quantity $kv_{F}\tau
=k\ell$ also appears in the loop integral. Since the Fermi velocity is
usually much larger than the sound velocity, $kv_{F}\gg \omega =v_{s}k$, the
limit $k\ell >1$ may often be reached in ultrasonic experiments \cite%
{Pippard} even though we always have $\omega \tau \ll 1$. Hence, expansion
of the result in powers of $k\ell$ may not be justified. In our case,
however, since we are considering sound propagation perpendicular to the
layers, $v_{F}$ is reduced by the anisotropy ratio $t_{\perp }/t_{||}$,
which can be as small as $10^{-4}$, so that it is justified in our case to
consider the limit $k\ell \ll 1$. One can thus expand in powers of $\omega
\tau $ and $k\ell $. The terms with odd powers of $\omega $ give corrections
to the sound attenuation while the terms proportional to even powers of $%
\omega $ contribute to the sound velocity. The first corrections that
involve $k$ appear to order $\left(\omega \tau \right) \left( k\ell \right)
^{2}$ because of the $p_{z}$ integral. In all that follows then, we set $k=0$
and expand in powers of $\left( \omega \tau \right) $.

From $\mathrm{Im}\left( \Pi ^{R}(\omega )\right) =-g^{2}\nu _{0}\omega \tau $
one can obtain the contribution to the imaginary part of the phonon
frequency \cite{AGD}
\begin{equation}
\gamma (\omega )=\frac{1}{2}\omega _{0}(k)\;\mathrm{Im}\left( \Pi^{R}(\omega
)\right) ,  \label{gamma}
\end{equation}
where is located the pole of Dyson's equation (\ref{Dyson_eq}). We assume
that the real part of the frequency at which the pole is located, $\omega $,
is close to the unperturbed frequency so that $\omega _{0}(k)\sim
\omega_{0}(k)^{2}/\omega \left( k\right) $. The power attenuation then reads
\begin{equation}
\alpha (\omega )=-2\gamma (\omega )/v_{s},  \label{alpha}
\end{equation}
where $v_{s}$ is the sound velocity. The sound attenuation increases with $
\tau $, as does conductivity.

The renormalization of the phonon frequency $\omega (k)$ is obtained from
the real part of phonon self-energy $\mathrm{Re}\left[ \Pi (\omega ,k)\right]%
^{R}=-g^{2} \nu_{0}\left( 1-\left( \omega \tau \right) ^{2}\right) $ using
\begin{equation}
\omega (k)= \omega_{0}(k)\sqrt{1+\mathrm{Re}\Pi ^{R}}.  \label{omaga_renorm}
\end{equation}

\section{Diagrams for fluctuation contribution to sound attenuation}

\begin{figure}[h]
\includegraphics{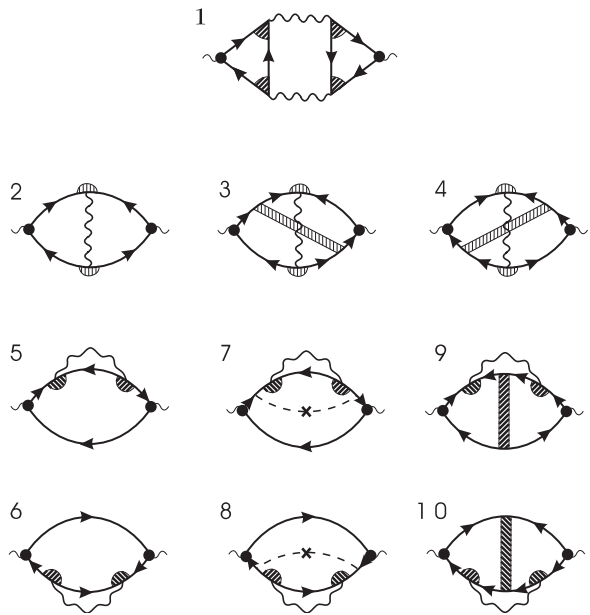} 
\caption{Feynman diagrams giving the leading-order corrections from the
fluctuations to the sound attenuation. Diagram 1 is of the Aslamazov-Larkin
(AL) type, diagrams 2-4 are of Maki-Thompson (MT) type, and the diagrams
5-10 are of the density of states (DOS) type. Solid lines are the normal
state Green functions, wavy lines the fluctuation propagators, shaded
semicircles and shaded rectangles the impurity ladder averagings, dashed
lines with cross the single impurity scatterings, and bold circles the
renormalized electron-phonon vertices. Adapted from Ref.\protect\cite{LV01}.}
\label{fl_diagrams}
\end{figure}

The preceding discussion shows that one can consider the same set of
diagrams, illustrated in Fig.\ \ref{fl_diagrams}, as for the conductivity\
\cite{LV01},\ \cite{Dorin93}, substituting $g\cos (p_{z}a)$ for the vector
vertices $ev_{\alpha }$ at the extreme left and right-hand sides.
Here, each wavy line in the figure corresponds to the fluctuation propagator
(Cooper ladder) $L(q,\Omega _{k})$ which, as $T\rightarrow T_{c}$, has the
form:
\begin{eqnarray}  \label{fluct_prop}
&&L(q,\Omega _{k})=-\left[ \nu _{0}\left( \epsilon +\psi \left( \frac{1}{2}+%
\frac{|\Omega _{k}|}{4\pi T}\right) -\psi \left( \frac{1}{2}\right) \right.
\right.  \label{Fluc_prop} \\
&&~~~~~~~-\left. \left. \frac{\langle ({\Delta \xi }\left( \mathbf{q,p}%
\right) _{\left| p\right| =p_{F}})^{2}\rangle _{F.S.}}{(4\pi T)^{2}}\psi
^{\prime \prime }\left( \frac{1}{2}+\frac{|\Omega _{k}|}{4\pi T}\right)
\right) \right] ^{-1},  \nonumber
\end{eqnarray}%
where $\epsilon \approx (T-T_{c})/T_{c}$, and $\psi (z)$ is the digamma
function. In the above Eq.\ (\ref{fluct_prop}), the angular average over the
Fermi surface with the spectrum in Eq.\ (\ref{spectrum}) reads:
\begin{eqnarray}
&&\langle ({\Delta \xi }\left( \mathbf{q,p}\right) _{\left| p\right|
=p_{F}})^{2}\rangle _{F.S.} \\
&&~~~~=\frac{1}{2}\left( \left( v_{F}q_{||}\right) ^{2}+\left( 4t_{\perp
}\sin \left( q_{z}a/2\right) \right) ^{2}\right) \equiv \tau ^{-1}\widehat{D}%
q^{2},  \nonumber
\end{eqnarray}%
where $\widehat{D}$ is a generalized diffusion operator.

Returning to Fig.\ \ref{fl_diagrams}, the shaded semicircles correspond to
vertex corrections from impurity averaging and are given by:
\begin{equation}  \label{lambda}
\lambda (\mathbf{q},\varepsilon _{1},\varepsilon _{2})=\frac{|\tilde{
\varepsilon}_{1}-\tilde{\varepsilon}_{2}|}{|\varepsilon _{1}-\varepsilon
_{2}|+\frac{\widehat{D}q^{2}}{\tau ^{2}|\tilde{\varepsilon}_{1}-\tilde{
\varepsilon}_{2}|^{2}}\Theta (-\varepsilon _{1}\varepsilon _{2})}.
\end{equation}
Diagrams 3, 4, 9 and 10 of Fig.\ \ref{fl_diagrams} contain the impurity
ladder in the particle-particle channel (shaded rectangle): 
\begin{eqnarray}
\Gamma _{c}(\mathbf{q};\varepsilon _{1},\varepsilon _{2}) &=&\frac{1}{2\pi
\tau \nu _{0}} \\
&\times &\left( 1-\frac{\Theta (-\varepsilon _{1}\varepsilon _{2})}{
|\varepsilon _{1}-\varepsilon _{2}|\tau }\left( 1-\frac{\widehat{D}q^{2}}{
\tau (\varepsilon _{1}-\varepsilon _{2})^{2}}\right) \right) ,  \nonumber
\end{eqnarray}
while diagrams 7 and 8 have a single impurity line, which corresponds to the
factor $1/(2\pi \tau \nu _{0})$. One can see that all other insertions of
impurity lines either lead to diagrams that vanish, or give negligible
corrections.

\section{Calculational details for the diagrams}

\subsection{MT and DOS diagrams}

We illustrate the procedure to evaluate the diagrams in Fig.\ \ref%
{fl_diagrams} by finding the expression for diagram 2 (Maki-Thompson type).
In analytical form it is given by the integral: 
\begin{eqnarray}
&&\Pi ^{(MT)}(\omega _{\nu },k)  \label{MT2_1} \\
&&~~~~~~~~=2g^{2}\ T\sum_{\Omega _{k}}\int \frac{d^{3}q}{(2\pi )^{3}}
L(q,\Omega _{k})K(q,\Omega _{k},\omega _{\nu }),  \nonumber
\end{eqnarray}
where 
\begin{widetext}
\begin{eqnarray}\label{MTint}
&&K(q,\Omega_k,\omega_\nu) = T\sum_{\eps_n} \lambda (q,\eps_n,  \Omega_k-\eps_n)\lambda
(q,\eps_n+\omega_\nu, \Omega_k-\eps_n-\omega_\nu)\\ &&~~~~~~\times
\int \frac{d^3p}{(2\pi)^3} \cos \left(p_z a\right) \cos \left((q_z-p_z)a \right)G(p,\eps_n)
G(p+k,\eps_n+\omega_\nu)G(q-p,\Omega_k-\eps_n)G(q-p-k,\Omega_k-\eps_n-\omega_\nu).\nonumber \end{eqnarray}
\end{widetext}
For simplicity, and as in experiment,\ \cite{usherb} we assume that the sound
propagates in the direction perpendicular to the conducting layers, so that $
\mathbf{k}$ has only one component: $\mathbf{k}=(0,k_{z})$. The most
singular contribution from the fluctuation propagator $L(\mathbf{q},\Omega
_{k})$ comes from $\mathbf{q}\rightarrow 0$. Therefore, one can neglect the $
\mathbf{q}$-dependence in the electron Green functions. We also set $\Omega
_{k}=0$ \cite{LV01} since superconducting fluctuations are important in the
renormalized classical regime where the characteristic frequency for the
pair propagator is less than temperature.

Performing the integration over $\mathbf{p}$ in Eq.\ (\ref{MTint}) first,
one has for $K(q,\omega _{\nu })\equiv K(q,0,\omega_\nu)$: 
\begin{eqnarray}
&&K(q,\omega _{\nu })=m\cos (q_{z}a)T \\
&&{\!\!\!\!\!\!\!\!\!}\times \sum_{\varepsilon _{n}}\frac{1}{\left(
2|\varepsilon _{n}+\omega _{\nu }|+\widehat{D}q^{2}\right) }\frac{1}{\left(
2|\varepsilon _{n}|+\widehat{D}q^{2}\right) }\frac{1}{|\widetilde{
\varepsilon _{n}+\omega _{\nu }}|+|\widetilde{\varepsilon }_{n}|}~.
\nonumber  \label{d22a}
\end{eqnarray}
that is then divided into two parts. Namely, the first part has the
summation in the interval $\varepsilon _{n}\in (-\infty ,-\omega _{\nu
})\cup \lbrack 0,\infty )$, while in the second one the summation is done in
the interval $\varepsilon _{n}\in \lbrack -\omega _{\nu },0)$,
\begin{eqnarray}
&&K(q,\omega _{\nu })  \label{d22b} \\
&&{\!\!\!\!\!\!\!\!\!}=2T\sum_{n=0}^{\infty }\frac{1}{\left( 2\varepsilon
_{n}+\omega _{\nu }+\widehat{D}q^{2}\right) }\frac{1}{\left( 2\varepsilon
_{n}+\widehat{D}q^{2}\right) }\frac{1}{2\varepsilon _{n}+\omega _{\nu }+\tau
^{-1}}~  \nonumber \\
&&+\frac{T}{\omega _{\nu }+\tau ^{-1}}\sum_{n=-\nu }^{-1}\frac{1}{\left(
2\varepsilon _{n}+2\omega _{\nu }+\widehat{D}q^{2}\right) }\frac{1}{\left(
-2\varepsilon _{n}+\widehat{D}q^{2}\right) }.  \nonumber
\end{eqnarray}

Each of these two contributions leads to a different temperature dependence.
The limits of summation in the second part of Eq.\ (\ref{d22b}) can be
obtained by taking into account the fact that frequencies $\varepsilon
_{n}=\pi T(2n+1)$ are of the fermionic type while the sound frequency $
\omega _{\nu }=2\pi T\nu $ is bosonic. The summation over $\varepsilon _{n}$
in the second part of Eq.\ (\ref{d22b}) results in:
\begin{equation}
-\frac{m\cos (q_{z}a)}{4\pi (1+\omega _{\nu }\tau )}\frac{\psi \left( \frac{
1 }{2}+\frac{\hat{D}q^{2}}{4\pi T}\right) -\psi \left( \frac{1}{2}+\frac{
\hat{D }q^{2}+2\omega _{\nu }}{4\pi T}\right) }{(\hat{D}q^{2}+\omega _{\nu
}) }.  \label{AMT}
\end{equation}
The analytical continuation of this sum involves some subtleties that are
discussed in the Appendix.

One can see that as $T\rightarrow T_{c}$, the main $q$-dependence in Eq.\ (%
\ref{MT2_1}) comes from the fluctuation propagator $L(q)$ and from the poles
of $K(q,\omega _{\nu })$. Therefore, it generally suffices to expand the
regular part of $K(q,\omega _{\nu })$ in powers of $q$ and to keep the very
first non-vanishing terms of the expansion.

The singular part of Eq.\ (\ref{AMT}) comes from taking the $q^{0}$-term of
the numerator together with the diffusion pole $1/(\hat{D}q^{2}+\omega _{\nu
})$ and analytically continuing them to real phonon frequencies. One arrives
at integrals of the type:
\begin{equation}
\int \frac{d^{3}q}{(2\pi )^{3}}\frac{\cos (q_{z}a)}{\left( \hat{D}%
q^{2}+1/\tau _{\phi }\right) \left( \epsilon +\eta q_{||}^{2}+r\sin
^{2}(q_{z}a/2)\right) }.  \label{AMT_qint}
\end{equation}%
where $r=16t_{\perp }^{2}\eta /v_{F}^{2}\ll 1$ is the anisotropy parameter
\cite{r}, and $\tau _{\phi }$ is the phase-breaking time, which provides the
convergence of the integrals over the pair momentum $q$ at small wave vector
$q$.\ \cite{Thompson} The coefficient $\eta $ has the meaning of the square of
the effective coherence length $\xi $ in the isotropic 2D case:\ \cite{LV01}
\begin{eqnarray}
&&\eta \equiv \xi ^{2}\left( T\tau \right) =-\frac{\left( \tau v_{F}\right)
^{2}}{2} \\
&&\times \left. \left[ \psi \left( \frac{1}{2}+\frac{1}{4\pi x}\right) -\psi
\left( \frac{1}{2}\right) -\frac{1}{4\pi x}\psi ^{^{\prime }}\left( \frac{1}{%
2}\right) \right] \right\vert _{x=T\tau }.  \nonumber
\end{eqnarray}%
At $T=T_{c}\,,$ the anisotropy parameter can be written as $r=4\xi _{\perp
}^{2}(0)/a^{2}$ where $\xi _{\perp }(0)$ is the Cooper pair size in the
perpendicular $\left( z\right) $ direction. %
We will also use the phase-breaking parameter $\gamma _{\phi }=2\eta
/(v_{F}^{2}\tau \tau _{\phi })$ instead of $\tau _{\phi }$.

The integral in Eq.\ (\ref{AMT_qint}) results in the so-called anomalous MT
contribution. Note however that while the leading term in the expansion of
the numerator of Eq.\ (\ref{AMT}) gave the above result, Eq.\ (\ref{AMT_qint}%
), the $q^{2}$-term of the expansion cancels the diffusion pole. This
``regular anomalous'' term, whose analog was overlooked in early work on
conductivity fluctuation corrections,\ \cite{VarlamovThanks} results in the
same type of integral as that given by the first part of Eq.\ (\ref{d22b})
with $q=0$:
\begin{equation}
\int \frac{d^{3}q}{(2\pi )^{3}}\frac{\cos (q_{z}a)}{\epsilon +\eta
q_{||}^{2}+r\sin ^{2}(q_{z}a/2)}.  \label{RMT_qint}
\end{equation}
Together, these two terms give so-called regular MT contribution, which
generally has weaker temperature dependence than the anomalous MT part.

The evaluation of the diagrams 5-8 of Fig.\ \ref{fl_diagrams} results in the
integrals of the type
\begin{equation}  \label{DOS_qint}
\int \frac{d^3 q}{(2\pi)^3} \frac{1}{\epsilon +\eta q_{||}^2 + r \sin^2(q_z
a/2)},
\end{equation}
which give the density of states part of the fluctuation corrections.

The remaining $\mathbf{q}$-integration in Eqs.\ (\ref{AMT_qint}), (\ref%
{RMT_qint}), and (\ref{DOS_qint}) is similar to that encountered for the
conductivity calculations and is described in much details in the review
\cite{LV01} and in Ref.\ \cite{Dorin93}. We should emphasize that as $
T\rightarrow T_{c}$, the main temperature dependence of the fluctuation
diagrams comes from these integrals. The remaining ``bubble", which consists
of Green functions and impurity lines, has a weaker dependence on $T\tau$.
It provides the coefficient in front of the function of $\epsilon $.

In the conductivity calculation, one can drop the diagrams containing the
impurity ladder (3, 4, 9 and 10 of Fig.\ \ref{fl_diagrams}). This is because
they contain the integration of one vector vertex with the Green function
triangle. Angular integration over $\mathbf{p}$ leaves only terms containing
powers of the small factor $(v_{F}k\tau )$. In the general case of sound
attenuation and propagation these diagrams would contribute, but for sound
propagating perpendicular to the layers the integration over the open Fermi
surface of the vertex $g\cos (p_{z}a)$ also leads to a vanishing
contribution for diagrams 3, 4, 9 and 10 of Fig.\ \ref{fl_diagrams} to
leading order in $k\ell $.

The small $\omega$ expansions of the DOS diagrams 5, 6, and of the regular
contributions to the MT diagram 2 each lead to zeroth-order terms, the
combination of which does not sum to zero. One notices the difference with
the conductivity calculation, where such a cancellation of zeroth-order
terms ensures that there is no anomalous diamagnetism above $T_{c}$. The DOS
diagrams 7, 8, and the anomalous part of the MT diagram 2 do not have
zeroth-order terms that would contribute to the leading renormalization of
the sound velocity.

\subsection{Aslamazov-Larkin diagram}

Finally, let us consider the Aslamazov-Larkin (AL) diagram 1, which gives
the main fluctuation correction to most of the transport phenomena. In the
case of sound propagation perpendicular to the layers, the analytical
expression for the AL diagram reads:
\begin{eqnarray}
&&\Pi ^{(AL)}(\omega _{\nu },k)=-2g^{2}\ T\sum_{\Omega _{k}}\int \frac{%
d^{3}q }{(2\pi )^{3}}B^{2}(q,\Omega _{k},\omega _{\nu })  \nonumber
\label{AL_1} \\
&&~~~~~~~~\times L(q,\Omega _{k})L(q,\Omega _{k}+\omega _{\nu }),
\end{eqnarray}
where 
\begin{eqnarray}
&&\!\!\!\!\!\!B(q,\Omega _{k},\omega _{\nu })=T\sum_{\varepsilon
_{n}}\lambda (q,\varepsilon _{n}+\omega _{\nu },-\varepsilon _{n})\lambda
(q,\varepsilon _{n},-\varepsilon _{n})  \label{AL_Bblock} \\
&&\!\!\!\!\!\!\times \int \frac{d^{3}p}{(2\pi )^{3}}\cos
(p_{z}a)G(p,\varepsilon _{n})G(p,\varepsilon _{n}+\omega _{\nu
})G(q-p,-\varepsilon _{n}),  \nonumber
\end{eqnarray}

In Eq.\ (\ref{AL_Bblock}), we have neglected the $\Omega _{k}$-dependence of
the electron Green functions. One can also set $\omega _{\nu }=0$ and expand
$G(q-p,-\varepsilon _{n})$ in Eq.\ (\ref{AL_Bblock}).\ \cite{Dorin93} For
sound propagating perpendicular to the planes, we can use $r \mathbf{\ll }1$ (See Eq.(\ref{r_limits})) to find the appropriate expansion
parameter for $G(q-p,-\varepsilon _{n})$ appearing in the integral, namely:
\[
\frac{\xi (\mathbf{q}-\mathbf{p})-\xi (\mathbf{p})}{\max \{\pi T,1/\tau \}}=
\frac{2t_{\perp }\left( \cos ((q_{z}-p_{z})a)-\cos (p_{z}a)\right) }{\max
\{\pi T,1/\tau \}}.
\]
After the angular integration in Eq.\ (\ref{AL_Bblock}), the leading
nonvanishing term reads:
\begin{equation}
B(q)=-\frac{4t_{\perp }\eta \nu _{0}(1-\cos (q_{z}a))}{v_{F}^{2}}.
\end{equation}
Following the Ref.\ \cite{Dorin93}, one can integrate Eq.(\ref{AL_1}), which
after the analytic continuation and expansion in powers of $\omega $,
results in the integral of the type:
\begin{equation}
\int \frac{d^{3}q}{(2\pi )^{3}}\frac{\sin ^{4}(q_{z}a/2)}{\left( \epsilon
+\eta q_{||}^{2}+r\sin ^{2}(q_{z}a/2)\right) ^{n+2}}  \label{AL_qint_2}
\end{equation}
in the $n$-th order of $\omega $-expansion. To leading orders in $\omega $,
the integrals in Eqs.\ (\ref{AL_qint_2}) are not divergent at $\epsilon
\rightarrow 0$. The resulting AL terms are finite and comparable with other
fluctuation contributions.

\section{Results \label{Sec.Results}}

In this section we collect the results for, in turn, the fluctuation
corrections to the sound attenuation and then to the sound velocity.

\subsection{Sound attenuation}

The odd powers of the $\omega $-expansion of the analytically continued
results for the diagrams of Fig.\ \ref{fl_diagrams} give the fluctuation
corrections to the imaginary part $\mathrm{Im}\left( \Pi ^{R}(\omega
)\right) $ of the retarded phonon self-energy. One can then use this result
to obtain the sound attenuation from the imaginary part of the pole of the
propagator, as in Eqs.\ (\ref{gamma}) and (\ref{alpha}).

All diagrams contribute to the polarization propagator to leading order in $
\omega $. The different contributions to the sound attenuation coefficient
are then extracted from the imaginary part of the retarded fluctuation
polarization operator. In the most general form, they can be written as:
\begin{equation}
\alpha ^{(\beta )}(T,\omega )=\frac{g^{2}\omega ^{2}\nu _{0}}{\varepsilon
_{F}v_{s}}\kappa _{att}^{(\beta )}(T\tau )f^{(\beta )}(\epsilon ,r,\gamma
_{\phi }),  \label{att_gen_form}
\end{equation}
where $\beta $ denotes the particular channel (DOS, rMT, aMT or AL). In Eq.\ (%
\ref{att_gen_form}), $f^{(\beta )}(\epsilon ,r,\gamma _{\phi })$ means the
function of temperature which comes from the integrals in Eqs.\ (\ref%
{AMT_qint}), (\ref{RMT_qint}), (\ref{DOS_qint}), and (\ref{AL_qint_2}),  and which also depends
on the material properties. The coefficient $\kappa _{att}^{(\beta )}(T\tau
) $ comes from the integration of Green functions and impurity blocks in the
expression of the type (\ref{MTint}). It has weaker temperature dependence
at $\epsilon \rightarrow 0$ than $f^{(\beta )}(\epsilon ,r,\gamma _{\phi })$%
, and it basically shows how the impurity concentration affects the result.

In the following, it will be more convenient to use the dimensionless
parameter $\tilde{\eta}(T\tau )=\eta /\left( \tau v_{F}\right) ^{2}$. We
list here the functions $f^{(\beta )}$, which also appear in the
conductivity calculation,\ \cite{Dorin93} and the coefficients $\kappa
_{att}^{(\beta )}$. For the reader's convenience, we also give the limiting
expressions of $f^{(\beta )}$ which depend on the relation between $\epsilon
$, $r$, and $\gamma _{\phi }$, where applicable.

\emph{DOS}:
\begin{eqnarray}  \label{DOS_f_alpha}
&&f^{(DOS)}(\epsilon,r)=\ln \left( \frac{2}{\sqrt{\epsilon }+\sqrt{ \epsilon
+r }}\right) \\
&&  \nonumber \\
&&~~~~~~~~~~\approx - \frac{1}{2}\left\{
\begin{array}{lll}
& \ln r\ ,\ ~~ & \epsilon \ll r, \\
&  &  \\
& \ln \epsilon\ ,\ ~~ & r \ll \epsilon.%
\end{array}
\right.  \nonumber
\end{eqnarray}
\begin{eqnarray}
&&\kappa _{att}^{(DOS)}(x)= \frac{ \psi^{^{\prime\prime}}\left( \frac{1}{2}
\right) -2\pi x \psi^{^{\prime}}\left( \frac{1}{2}+\frac{1}{4\pi x }\right)}{
8\pi^2 x \tilde{\eta}(x)}  \nonumber \\
&&~~~~~~~~~~~~~~\approx \left\{
\begin{array}{lll}
& -{\ \mbox{\Large $\frac{28 \zeta(3)}{\pi^3}$}}, & ~~x\ll 1, \\
&  &  \\
& -{\ \mbox{\Large $\frac{4\pi^3 x^2 }{7 \zeta(3)}$}}, & ~~x\gg 1.%
\end{array}
\right.
\end{eqnarray}

\emph{Regular part of MT}:
\begin{eqnarray}  \label{rMT_f_alpha}
&&f^{(rMT)}(\epsilon,r)= \frac{\left(\sqrt{\epsilon}-\sqrt{\epsilon+r}
\right)^2 }{r} \\
&&  \nonumber \\
&&~~~~~~\approx \left\{
\begin{array}{lll}
& 1\ ,\ ~~ & \epsilon \ll r, \\
&  &  \\
& {\mbox{\Large $\frac{r}{4 \epsilon }$}}\ ,\ ~~ & r \ll \epsilon.%
\end{array}
\right.  \nonumber
\end{eqnarray}
\begin{eqnarray}
&&\kappa _{att}^{(rMT)}(x)= -\frac{\pi^3 x -2 \pi x \psi^{^{\prime}}\left(
\frac{1}{2}+\frac{1}{4\pi x}\right) + \psi^{^{\prime\prime}}\left( \frac{1}{
2 }\right)}{16 \pi^2 x \tilde{\eta }(x)}  \nonumber \\
&&~~~~~~~~~~~~~~\approx \left\{
\begin{array}{lll}
& {\mbox{\Large $\frac{14 \zeta(3)}{\pi^3}$}}, & ~~x\ll 1, \\
&  &  \\
& {2 x}, & ~~x\gg 1.%
\end{array}
\right.
\end{eqnarray}

\emph{Anomalous part of MT}:
\begin{eqnarray}  \label{aMT_f_alpha}
&&f^{(aMT)}(\epsilon,r,\gamma_\phi)=\frac{1}{r} \left(-1+\frac{
\epsilon+r+\gamma_\phi}{\sqrt{\gamma_\phi(\gamma_\phi+r)}+ \sqrt{
\epsilon(\epsilon+r)}} \right)  \nonumber \\
&&  \nonumber \\
&&~~~~~~~~~~~~~~~~~\approx \left\{
\begin{array}{lll}
& {\mbox{\Large $ \frac{1}{\sqrt{r \gamma _{\phi }}}$}}\ ,\ ~~ & \epsilon
\ll \gamma _{\phi }\ll r, \\
&  &  \\
& {\mbox{\Large $\frac{1}{\sqrt{r \epsilon }}$}}\ ,\ ~~ & \gamma _{\phi }\ll
\epsilon \ll r, \\
&  &  \\
& {\mbox{\Large $\frac{1}{ 2\epsilon}$}}\ ,\ ~~ & \gamma _{\phi }\ll r\ll
\epsilon, \\
&  &  \\
& {\mbox{\Large $\frac{1}{ 2\gamma_\phi}$}}\ ,\ ~~ & \epsilon\ll r\ll \gamma
_{\phi }, \\
&  &  \\
& {\mbox{\Large $\frac{r}{ 8 \gamma_\phi \epsilon}$}}\ ,\ ~~ & r\ll {%
\mbox{min}}\{\gamma_\phi,\epsilon\}.%
\end{array}
\right.
\end{eqnarray}
\begin{equation}
\kappa _{att}^{(aMT)} = - \frac{\pi}{4}.
\end{equation}

\emph{AL}:
\begin{eqnarray}  \label{AL_f_alpha}
&&f^{(AL,\alpha)}(\epsilon,r)=\frac{1}{r} \left(1-\sqrt{\frac{ \epsilon}{
\epsilon+r}} \left(1+\frac{r}{2(\epsilon+r)}\right) \right)  \nonumber \\
&&  \nonumber \\
&&~~~~~~\approx \left\{
\begin{array}{lll}
& {\mbox{\Large $\frac{1}{r}$}}\ ,~~ & \epsilon \ll r, \\
&  &  \\
& {\mbox{\Large $\frac{3 r}{8 \epsilon^2}$}}\ ,\ ~~ & r \ll \epsilon.%
\end{array}
\right.
\end{eqnarray}
\begin{equation}
\kappa _{att}^{(AL)} = \frac{\pi}{8}.
\end{equation}

The fluctuation terms scale like $\omega ^{2}$ just as in the corresponding
normal metal case without fluctuation correction.

\subsection{Sound velocity}

The renormalization of the phonon frequency $\omega (k)$ obtained from Eq.\ (%
\ref{omaga_renorm}) provides the fluctuation corrections to the sound
velocity when we use $\omega (k)=v_{s}k$. It should be emphasized that there
is no anomalous MT term in leading (zeroth) order of $\omega $, while to
higher order of $\omega$ all types of temperature dependences appear. To
leading order $\omega ^{0}$ the non-zero terms give, for $\Delta \omega
/\omega =\left( \omega -\omega _{0}\right) /\omega _{0}$, the expressions in
a form:
\begin{eqnarray}
&&\frac{\Delta \omega ^{(\beta )}(T,\omega )}{\omega }=\frac{\Delta
v_{s}^{(\beta )}\left( T,\omega \right) }{v_{s}}  \label{v_gen_form} \\
&&~~~~~~~~~~~~=\frac{g^{2}T\nu _{0}}{\varepsilon _{F}}\kappa _{v}^{(\beta
)}f^{(\beta )}(\epsilon ,r,\gamma _{\phi }),  \nonumber
\end{eqnarray}
For DOS, rMT and aMT terms, the temperature functions $f^{(\beta )}(\epsilon
,r,\gamma _{\phi })$ are given by Eqs.\ (\ref{DOS_f_alpha}), (\ref%
{rMT_f_alpha}) and (\ref{aMT_f_alpha}), respectively. The temperature
function $f^{(AL,v)}(\epsilon ,r)$ for the AL diagram is given by the Eq.\ (%
\ref{AL_f_v}), which is different from that for attenuation (\ref{AL_f_alpha}%
). We then obtain for the coefficients $\kappa_{v}^{(\beta )}$:

\emph{DOS}:
\begin{eqnarray}
\kappa_{v}^{(DOS)}=1,
\end{eqnarray}

\emph{Regular part of MT}:
\begin{eqnarray}
\kappa_{v}^{(rMT)}=-\frac{1}{2},
\end{eqnarray}


\emph{Anomalous part of MT}:
\begin{eqnarray}
\kappa_{v}^{(aMT)}=0,
\end{eqnarray}

\emph{AL}:
\begin{eqnarray}  \label{AL_f_v}
&&\kappa _{v}^{(AL)} f^{(AL,v)}(\epsilon,r) \\
&&~~~~~~~=-\left(1+ \frac{2 \epsilon}{r} \left(-1+\sqrt{\frac{ \epsilon}{
\epsilon+r}}\right)\right)  \nonumber \\
&&  \nonumber \\
&&~~~~~~~~~~~\approx - \left\{
\begin{array}{lll}
& 1\ ,~~ & \epsilon \ll r, \\
&  &  \\
& {\mbox{\Large $\frac{3 r}{4 \epsilon}$}}\ ,\ ~~ & r \ll \epsilon.%
\end{array}
\right.  \nonumber
\end{eqnarray}

\section{Discussion\label{Sec_Discussion}}

Comparison of the normal state results without fluctuation corrections found
in Sec.\ \ref{Normal} with Eqs.\ (\ref{att_gen_form}) and (\ref{v_gen_form})
above show that the fluctuation corrections are smaller than the normal
state contribution by a factor of $T/\varepsilon _{F}$. For layered
organics, the Fermi surface parameters of Ref.\ \cite{Dressel} lead to $%
T/\varepsilon _{F}\sim 10^{-2}$. The temperature functions $f^{(\beta )}$
increase this ratio at $T\rightarrow T_{c}$, thus making the fluctuation
corrections experimentally measurable.

It can be seen from the results of previous section that it is the $\kappa
^{(\beta )}(T\tau )$ coefficients that determine the sign of the fluctuation
sound attenuation and sound velocity. The coefficients for sound attenuation
$\kappa _{att}^{(\beta )}(T\tau )$ in Eq.\ (\ref{att_gen_form}) have weaker
temperature dependence than that of the functions $f^{\beta }$. One can see
that these coefficients in all fluctuation diagrams are finite as $(T\tau
)\rightarrow 0$, and that they are of the same order of magnitude in this
limit. In the opposite (clean) limit $(T\tau )\gg 1$, these coefficients
increase as a power law of $(T\tau )$. We recall that there is an analogous
formal divergence of the DOS and MT coefficients \cite{Dorin93} in the
fluctuation conductivity but in that case they cancel each other in the
limit $(T\tau )\gg 1$. In our case, the formal extension of our results to
the limit $(T\tau )\gg 1/\sqrt{\epsilon }$ would be incorrect because the
local approximation for the fluctuation propagator (\ref{fluct_prop}) and
the impurity vertex (\ref{lambda}) have a natural limit of validity $(T\tau
)\ll 1/\sqrt{\epsilon }$ (the so-called local clean case).\ \cite{LSV00}

In contrast to the coefficients for sound attenuation $\kappa _{att}^{(\beta
)}(T\tau )$, the coefficients in the case of sound velocity, $\kappa
_{v}^{(\beta )}$ in Eq.\ (\ref{v_gen_form}), do not depend on $T\tau $, thus
simplifying the estimation of the sound velocity corrections in particular
materials.

The overall magnitude of the fluctuation correction is given mainly by the
temperature functions $f^{(\beta )}$. These functions depend on the
anisotropy parameter $r$ and the phase-breaking time $\tau _{\phi }$, which
are defined by the intrinsic material properties. The temperature function
corresponding to a given type of diagram is identical for sound attenuation
and velocity, except for the AL contribution which has temperature
dependences that differ. In general, all the fluctuation terms remain finite
as $T\rightarrow T_{c}$. The asymptotics of the corresponding temperature functions have been given in the preceding section and they were also
discussed elsewhere.\ \cite{Dorin93}

\begin{figure}[h]
\includegraphics[width=8.0cm]{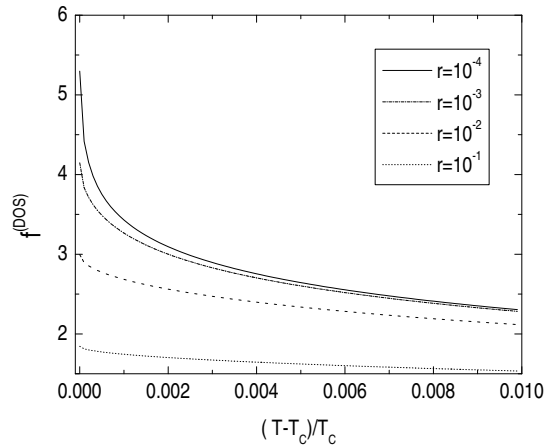}
\caption{The temperature dependence of $f^{(DOS)}$ at various values of
anisotropy parameter $r=10^{-1}\ldots 10^{-4}$.}
\label{DOS_fig}
\end{figure}
\begin{figure}[h]
\includegraphics[width=8.0cm]{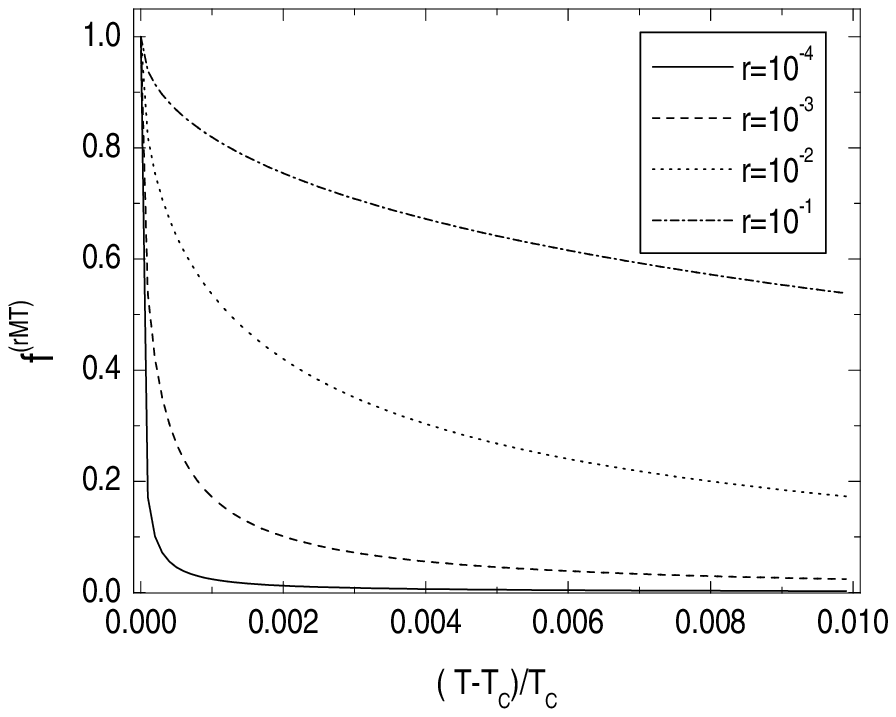}
\caption{The temperature dependence of $f^{(rMT)}$ at various values of
anisotropy parameter $r=10^{-1}\ldots 10^{-4}$.}
\label{RMT_fig}
\end{figure}
\begin{figure}[h]
\includegraphics[width=8.0cm]{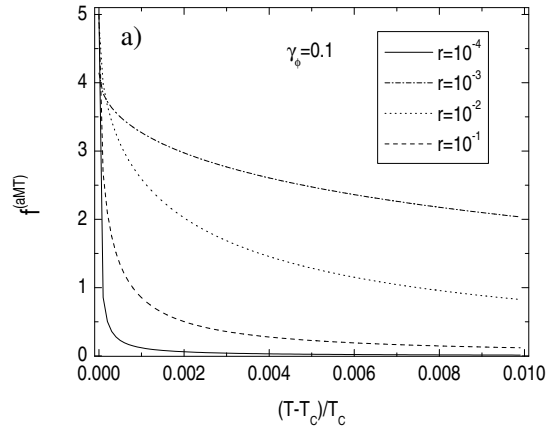} %
\includegraphics[width=8.0cm]{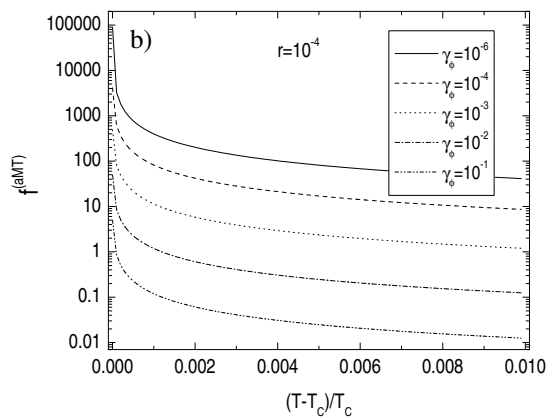}
\caption{The temperature dependence of $f^{(aMT)}$ a) at fixed realistic $%
\protect\gamma _{\protect\phi} =0.1$ and at various values of anisotropy
parameter $r=10^{-1}\ldots 10^{-4}$, b) at most realistic for the $\protect%
\kappa $-(ET)$_{2}$ family compounds $r=10^{-4}$ and at various $\protect%
\gamma _{\protect\phi} =10^{-1}\ldots 10^{-6}$.}
\label{AMT_fig}
\end{figure}
\begin{figure}[h]
\includegraphics[width=8.0cm]{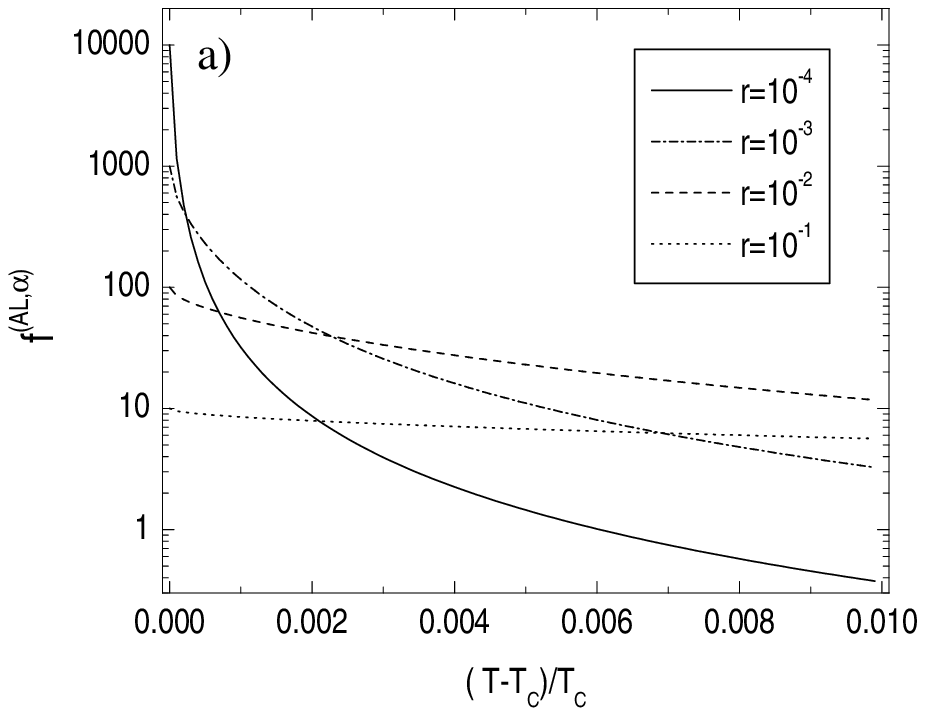} %
\includegraphics[width=8.0cm]{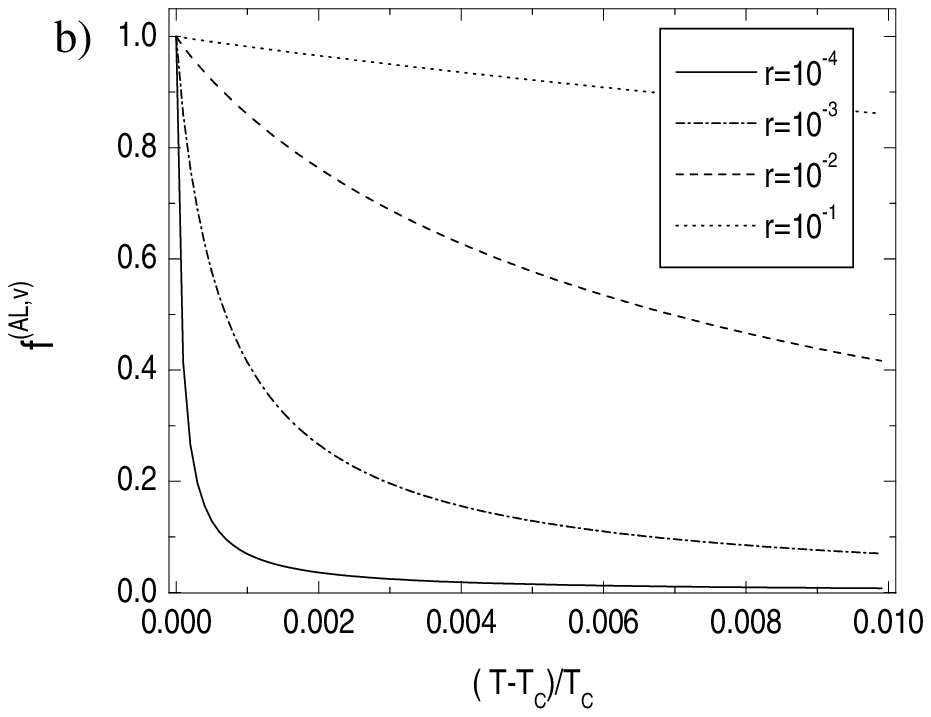}
\caption{The temperature dependence of AL contributions a) to the sound
attenuation $f^{(AL,\protect\alpha )}$, b) to the sound velocity $f^{(AL,v)}$%
, at various values of anisotropy parameter $r=10^{-1}\ldots 10^{-4}$.}
\label{AL_fig}
\end{figure}

The functions $f^{\left( \beta \right) }$ also describe the crossover
between two and three dimensions, where by three-dimensions we mean that the
superconducting correlation length is much larger than the interlayer
spacing. The two-dimensional behavior corresponds to $\varepsilon \gg r$ and
the three-dimensional behavior to the opposite limit $r\gg \varepsilon $.
The crossover occurs at $\varepsilon \sim r$, namely when the superconducting
correlation length $\xi _{\perp }(0)\left\vert \varepsilon \right\vert
^{-1/2}$ in the perpendicular direction becomes of the order of the
interplane spacing.

In order to compare the contributions from different diagrams, we consider a
range of parameters which is realistic for layered organic superconductors.
The estimation for the anisotropy $t_{||}/t_{\perp }\approx 4000$, where $%
t_{||}$ is the intralayer transfer integral, is in agreement with the
magnetoresistance data for $\kappa $-(ET)$_{2}$Cu(NCS)$_{2}$.\ \cite%
{Singleton} This gives an anisotropy parameter of the order of $r{\ \gtrsim
\ }10^{-4}$.

Concerning the scattering time, its estimate \cite{Singleton_tau} from
Shubnikov-de Haas experiments in $\kappa $-(ET)$_{2}$ materials gives $\tau
\approx 3\;ps$, which leads to $T\tau \sim 1$ at $T\geq T_{c}$.

The phase-breaking time $\tau _{\phi }$ can be estimated from the
experimental data on fluctuation contribution to transport coefficients
using the appropriate expression for the MT diagram. The analysis of the
magnetoresistance data on YBCO in Ref. \cite{Larkin_tau}, which included AL and MT
contributions, resulted in values $\tau _{\phi }\sim 1/T$. Later on, the authors of Ref.
\cite{Dorin93} claimed that neglecting the orbital pair breaking effects while leaving the Zeeman
contribution to the MT diagram in Ref. \cite{Larkin_tau} can be incorrect. Hence, they extended the
range of the realistic values of $\tau _{\phi }$ to $\tau _{\phi }{\ \lower-1.2pt
\vbox{\hbox{\rlap{$
          <$}\lower5pt\vbox{\hbox{$\sim$}}}}\ }10/T$ with $\tau _{\phi }\sim
100/T$ being still acceptable for the analysis.\ \cite{Dorin93} A similar
estimate was chosen by the authors of Ref.\ \cite{Sauls_tau} to predict the
magnitude of the fluctuation corrections to the nuclear spin-lattice
relaxation rate and the NMR Knight shift in high-$T_{c}$ cuprates. To the
best of our knowledge, no relevant analysis in layered organic materials
have been reported to date. However, it is known that the phase-breaking
time should satisfy the relation $\tau _{\phi }>\tau $. Without loss of
generality then, we assume that pair breaking by inelastic scattering is
weak enough and we consider $\gamma _{\phi }\ \sim 10^{-3}...10^{-2}$.

The aMT contribution vanishes to leading order, $\omega ^{0}$, for sound
velocity. With the above estimation for the anisotropy parameter, one can
see that the AL part as well as the regular MT part give the least important
non-vanishing contributions to sound velocity (order $\omega ^{0}$). Indeed,
$\lim_{\varepsilon \rightarrow 0}f^{(rMT)}(\varepsilon ,r)={\
\lim_{\varepsilon \rightarrow 0}}f{^{(AL,v)}(\varepsilon ,r)=1}$ at any $r$,
while for the DOS contribution, $\lim_{\varepsilon \rightarrow
0}f^{(DOS)}(\varepsilon ,r)=(1/2)\ln (4/r)\approx 5$ at $r\sim 10^{-4} $.

For sound attenuation the contribution of the AL term can be significant
near $T_{c}$, by contrast to its contribution to the sound velocity, since $%
\lim_{\varepsilon \rightarrow 0}f^{(AL,\alpha )}(\varepsilon ,r)=1/r$.
However, it decays much faster than all other fluctuation terms above $T_{c}$%
: $f^{(AL,\alpha )}(\epsilon =0.002)/f^{(AL,\alpha )}(\epsilon =0.01)\simeq
23$ with $\lim_{r/\varepsilon \rightarrow 0}f^{(AL,\alpha )}(\varepsilon
)\sim 1/\epsilon ^{2}$. At the same time, at $\gamma _{\phi }\sim
10^{-3}...10^{-2}$ and $r\sim 10^{-4}$, we see that the anomalous MT
temperature function is large, $f^{(aMT)}(\epsilon =0)\approx 500$ at $%
\gamma _{\phi }\sim 10^{-3}$ and $f^{(aMT)}(\epsilon =0)\approx 50$ at $%
\gamma _{\phi }\sim 10^{-2}$. Thus, at some $\gamma _{\phi }$, the aMT can
be much larger than all other contributions. However, while it is quite
large at $\epsilon =0$, the aMT part decreases with increasing $T$ faster
than the DOS contribution. Indeed, at $r=10^{-4}$, the ratio $%
f^{(DOS)}(\epsilon =0.002)/f^{(DOS)}(\epsilon =0.01)\simeq 1.3$ shows that
the DOS contribution has not changed much in that temperature range while
the corresponding ratio for aMT is $f^{(aMT)}(\epsilon
=0.002)/f^{(aMT)}(\epsilon =0.01)\simeq 5$ for the same $r$ and in the
relevant range of $\gamma _{\phi }$. In other words, the logarithmic
asymptotics (\ref{DOS_f_alpha}) of the DOS diagram at $r/\epsilon
\rightarrow 0$ wins over the power-law dependencies of the aMT term (\ref%
{aMT_f_alpha}).

One can see that next order terms in the frequency expansion for the sound
velocity and sound attenuation would contain the factor $(\omega /T)^{2n}$,
where $n=1,2,\ldots $. The typical ultrasound frequency is of the order of $
\omega \sim 100$ $\mathrm{MHz}$. At $T\gtrsim T_{c}\simeq 10\;\mathrm{K}$,
the relevant frequency to temperature ratio is thus of the order $(\omega
/T)^{2}\sim 10^{-7}$. One can check, for example, that in order to
compensate for such a small factor, the parameter $\gamma _{\phi }$ in the
anomalous MT term at $r\sim 10^{-4}$ should be as small as $10^{-10}$,
corresponding to unreasonably large $\tau _{\phi }\sim 10^{8}\tau $. We
conclude that, in our model, the sound velocity should be increased by the
fluctuation corrections given by the DOS term.

Finally we would like to comment on the applicability of our model with regard to the character of  interlayer electron transport in  organic
conductors, a subject which is still hotly debated. The interlayer tunneling of electrons  can be coherent or
incoherent (see Refs. \cite{Timusk_organics,MosesMcKenzie} and references therein). Following the
terminology of Ref. \cite{MosesMcKenzie}, we should distinguish between two limiting cases of
incoherent interlayer transport, strongly and weakly incoherent. First, in the strongly incoherent case the intralayer electron
momentum is not conserved in the tunneling processes between   adjacent
layers because the tunneling can be accompanied by strong elastic or
inelastic processes. In this case the motion of electrons between the layers
is diffusive and the electron band states in the $z$ direction and the
corresponding Fermi surface cannot be defined. Our model is invalid in that
limit. It is valid, however, in the second case, namely for weakly
incoherent tunneling, that occurs when $\tau t_{\perp }\ll 1.$ In that case
the intralayer electron momentum is conserved in the tunneling process and
the electron wave function  in adjacent layers has some overlap, but there
are many in-plane collisions between tunneling events. Our approach is also valid obviously in the
coherent case where $\tau t_{\perp }\gg 1$.

It is clear from  our results  that we can also consider two other
limiting cases, namely the clean limit ($T\tau \gg 1$) and the dirty limit ($%
T\tau \ll 1$). The latter always corresponds to weakly incoherent tunneling
while both coherent and weakly incoherent tunneling can occur in the former.
This can be explained as follows. We have assumed throughout that $r\ll 1$.
That parameter takes the following limiting forms:\ \cite{Dorin93}
\begin{equation}
r=\frac{4\xi _{\perp }^{2}(0)}{a^{2}}\sim \left\{
\begin{array}{cc}
{\mbox {\Large $\frac{t_{\perp }^{2}\tau }{T}$ }},\; & T\tau \ll 1 \\
&\\
{\mbox {\Large $\frac{t_{\perp }^{2}}{T^2}$ }},\; & T\tau \gg 1%
\end{array}%
\right.   \label{r_limits}
\end{equation}%
where at $T=T_{c}$,  $\xi _{\perp }(0)$ is the size of the Cooper pair
perpendicular to the plane. The conditions $r\ll 1$ $,$ and $T\tau \ll 1$,
corresponding to dirty case, are consistent only with the weakly incoherent
tunneling since they lead to $\left( \tau t_{\perp }\right) ^{2}\ll T\tau
\ll 1$. On the other hand, for $T\tau \gg 1$ the condition $r\ll 1$ is
consistent with both coherent and weakly incoherent tunneling. This explains why in the preceeding discussion  we distinguished only between the clean and dirty limits. Note that the crossover from clean to dirty limit occurs at $v_F/T \sim v_F\tau$, namely when the elastic mean free path becomes smaller than the thermal de Broglie wavelength.

In summary, the superconducting fluctuation corrections to the sound
attenuation, in a realistic range of parameters, are given by the sum of the
DOS, anomalous MT{\ and AL} terms, which decrease the normal state
attenuation. In contrast, to leading order in $\omega $, the corrections to
the sound velocity are given by DOS diagrams, because the expansion for the
anomalous MT diagram begins at order{\ $\omega $ while the AL diagram is
small to this order.}

\section{Conclusion}

We calculated, in the hydrodynamic limit $\omega \tau \ll 1$ and $k\ell \ll
1 $, the effect of superconducting fluctuations on longitudinal sound
propagating perpendicular to the layers of quasi two-dimensional systems
above $T_{c}$. This corresponds to the experimentally realizable situation.
The detailed results for the sound attenuation coefficient $\alpha $ and the
sound velocity renormalization $\Delta v_{s}/v_{s}$ are summarized in
section \ref{Sec.Results}. {\ In short, there are four types of temperature
dependences coming respectively from a) the density of states diagrams, b)
the regular contribution of the Maki-Thompson diagram, c) the anomalous
contribution of the Maki-Thompson diagram and d) from AL diagram.} These
contributions have different signs and their magnitudes strongly depend on
material properties such as anisotropy and phase-breaking processes. {For
longitudinal sound propagating perpendicular to the layers, the open Fermi
surface condition leads to a suppression of the divergent contributions to
leading order,} so that fluctuation corrections to sound attenuation are
expected to generally be much smaller than the fluctuation corrections to
conductivity.

The leading temperature dependences with respect to $T_{c}$ of the
fluctuation corrections to the sound attenuation have the same $\omega ^{2}$
scaling with frequency as in the impure (dirty) metal without fluctuation
correction. To this order in $\omega $, all {four} types of temperature
contributions are present. As we move close to $T_{c}$, the fluctuation
contribution decreases the sound attenuation below its normal state level,
at least for realistic values of the material parameters.

The leading fluctuation correction to the sound velocity does not depend on $
\omega $, just like the normal state term. To this order, there are only
three types of temperature dependences coming from the density of states
diagram, {AL diagram} and from the regular contribution of the Maki-Thompson
diagram. The overall fluctuation correction in realistic material situation
should be dominated by the density of states contribution and should
increase the sound velocity as we approach $T_{c}$. The anomalous
contribution of the Maki-Thompson diagram is proportional to $\left( \omega
/T\right) ^{2} $ and generally can be neglected unless phase breaking is
extremely low. Further details of our calculations may be found in Ref.\
\cite{web}.

It should be noted that our results on sound attenuation are in qualitative
agreement with the preliminary experimental data.\ \cite{KFrikash} These
experiments are currently in progress. The detailed analysis including the
comparison with the present theory will be published elsewhere.

To make actual comparisons with experiment, one may need to consider several
other physical effects. For example, strongly incoherent interplane
tunneling \cite{MosesMcKenzie} and $d$-wave superconductivity \cite{Mayaffre95} may have to be
considered. In addition, one may need to include in the fluctuation
propagator Eq.\ (\ref{Fluc_prop}) mode-coupling terms that will allow to go
beyond mean-field theory in the description of the crossover from two to
three dimensions.

\begin{acknowledgments}
The authors would like to thank A.A.~Varlamov, K.V.~Samokhin, K.~Maki and
L.V.~Bulaevskii for valuable discussions and especially J.-S.~Landry for
performing preliminary calculations. We are thankful to M.~Poirier,
D.~Fournier and J.~Singleton for information on recent experiments in
organic superconductors. A.-M.S.T. would like to acknowledge the hospitality
of Universit\'{e} de Provence and of Yale University where part of this work
was performed and, in the latter case, supported under NSF Grant-0342157. The present work was
supported by the Natural Sciences and Engineering Research Council (NSERC)
of Canada, the Fonds de la Recherche sur la Nature et les Technologies of
the Qu\'{e}bec government, and the Tier I Canada Research Chair Program
(A.-M.S.T.).
\end{acknowledgments}

\appendix

\section{On the analytical continuation of the digamma function of complex
argument}

We shall discuss an important mathematical issue occurring in the evaluation
of the anomalous MT contribution (\ref{d22b}) and its further analytic
continuation. Namely, let us look at the typical sum over the fermionic
Matsubara frequencies:
\begin{equation}
T\sum_{n=0}^{\nu -1}\frac{1}{\varepsilon _{n}}.  \label{sum_example}
\end{equation}
The expression (\ref{sum_example}) appears if one decomposes into partial
fractions the initial sum over $\varepsilon _{n}$ in the anomalous part of
the Maki-Thompson diagram and neglects the small terms $\sim \hat{D}q^{2}$
in them. One can easily evaluate the sum (\ref{sum_example}) using the
well-known representation of the digamma function:
\begin{equation}
T\sum_{n=0}^{\nu -1}\frac{1}{\varepsilon _{n}}=\frac{1}{2\pi }\left\{ \psi
\left( \frac{1}{2}+\nu \right) +2\ln 2+\gamma \right\} ,  \label{Plus}
\end{equation}
where $\gamma =0.577216$ is the Euler constant. Here $\nu $ comes from the
external bosonic Matsubara frequency $\omega _{\nu }=2\pi T\nu $, with $\nu $
an integer. At the same time, it is easy to see that the straightforward
change of variable $n=-n^{\prime }-1$ transforms the initial expression (\ref%
{sum_example}) in a way that one obtains the alternate result:
\begin{eqnarray}
&&T\sum_{n=0}^{\nu -1}\frac{1}{\varepsilon _{n}}\equiv -T\sum_{n^{\prime
}=-\nu }^{-1}\frac{1}{\varepsilon _{n}^{\prime }}  \label{Minus} \\
&&~~~~~~~~~~~=\frac{1}{2\pi }\left\{ \psi \left( \frac{1}{2}-\nu \right)
+2\ln 2+\gamma \right\}.  \nonumber
\end{eqnarray}
As long as we deal with the Matsubara frequencies $\omega _{\nu }=2\pi T\nu $
these two results Eq.\ (\ref{Plus}) and Eq.\ (\ref{Minus}) are equivalent
because of the relation
\begin{equation}
\psi \left( \frac{1}{2}-z\right) =\psi \left( \frac{1}{2}+z\right) +\pi \tan
(\pi z),  \label{tan}
\end{equation}
and the fact that $\tan (\pi z)=0$ when $z=\nu $ is an integer. Therefore,
at this stage we are free to choose either Eq.\ (\ref{Plus}) or Eq.\ (\ref%
{Minus}) as the result of the summation. The distinction appears when one
tries to do the analytic continuation to real frequencies $i\omega _{\nu
}\rightarrow \omega +i\eta $. After analytic continuation, Eqs.\ (\ref{Plus}%
) and (\ref{Minus}), which we obtained from the same diagram, are not equal
to each other anymore. Indeed, for non-integer $\nu $, $\psi \left( \frac{1}{%
2} -z\right) =\psi \left( \frac{1}{2}+z\right) $ is not true anymore, as one
can see from Eq.\ (\ref{tan}).

It is crucial to choose either Eq.\ (\ref{Plus}) or Eq.\ (\ref{Minus}) as
the result of the summation before we do the analytic continuation. The
choice is determined by whether we want the advanced or the retarded
response function. Indeed, consider Eqs.\ (\ref{Plus}) and (\ref{Minus}) as
general response functions $\chi ^{M}(\omega _{\nu })=\chi (z)|_{z=i\omega
_{\nu }}\varpropto \psi \left( \frac{1}{2}\pm \nu \right) $ on the Matsubara
frequency domain $\omega _{\nu }$. If we then analytically continue the
argument $z=i\omega _{\nu }\rightarrow \omega +i\eta $ of $\psi \left(
1/2-iz/(2\pi T)\right) $, the poles of resulting function are all located
below the real axis of the complex plane thus giving us the retarded
function $\chi ^{R}(\omega )$. On the other hand, the analytic continuation $
z=i\omega _{\nu }\rightarrow \omega -i\eta $ of the argument of $\psi \left(
1/2+iz/(2\pi T)\right) $ gives us the advanced response because all its
poles are located in the upper half-plane.

We thus conclude that if the digamma function appears in our calculations,
we are free to use either Eqs.\ (\ref{Plus}) or (\ref{Minus}) because they
are equivalent, as long as we deal with Matsubara frequencies. However,
before doing the analytic continuation, we have to decide whether we need
the advanced or the retarded function. If we want the retarded function, we
use the rule given by Eq.\ (\ref{Plus}) and deal only with digamma functions
of the type $\psi (1/2+\nu )$. Otherwise, use $\psi (1/2-\nu )$ to obtain
the advanced response.

\end{document}